\documentclass[10pt,twocolumn]{article}

\usepackage[utf8]{inputenc}
\usepackage[T1]{fontenc}

\usepackage{graphicx}
\usepackage{pdfpages}

\usepackage{lipsum}
\usepackage{color}

\usepackage{authblk}
\usepackage[top=20mm, bottom=20mm, right=15mm, left=15mm]{geometry}
\usepackage{mathtools, cuted}

\usepackage{bm}
\usepackage{amsfonts}
\usepackage{amsmath}
\usepackage{amssymb}
\usepackage{mathrsfs} % cursive letters
\usepackage{siunitx}

\title{Evaporation of liquid coating a fiber}
\author[1]{M. Corpart}
\author[2]{J. Dervaux}
\author[1]{C. Poulard}
\author[1]{F. Restagno}
\author[1]{F. Boulogne}
\affil[1]{Université Paris-Saclay, CNRS, Laboratoire de Physique des Solides, 91405, Orsay, France.}
\affil[2]{Université  de Paris,  Laboratoire  Mati\`ere  et  Syst\`emes  Complexes, UMR  7057  CNRS,  F-75013  Paris,  France}

\date{\today}
\begin{document}

\twocolumn[
    \begin{@twocolumnfalse}
        \maketitle
        \begin{abstract}
We investigate theoretically and numerically the diffusion-limited evaporation of a liquid deposited on a fiber in two configurations: a sleeve and a axisymmetric barrel-shaped droplet.
For a sleeve, the local flux depends on both the aspect ratio and the smallest length of the problem.
By  using analytical calculations and 3D finite elements simulations, we predict a divergence of this flux further localized at the edge as the aspect ratio increases.
The evaporation of axisymmetric drops on a fiber is studied with numerical simulations.
For sufficiently large volumes, we evidence that the evaporation rate is almost independent of the wetting properties of the liquid, even for small contact angles, and that the droplets evaporate as spheres of the same volume.
        \end{abstract}
    \end{@twocolumnfalse}
]

%%%%%%%%%%%%%%%%%%%%%%%%%%%%%%
%
% INTRODUCTION
%
%%%%%%%%%%%%%%%%%%%%%%%%%%%%%%
\section{Introduction}

The coating of liquid on a solid is a common operation in everyday life as well as in industrial processes (\textit{e.g.} lubrication, painting).
The first theoretical description of the deposited layer has been made almost simultaneously by Landau and Levich \cite{Landau1942} and Derjaguin \cite{Derjaguin1943}.
The coated thickness $h$ is found to be proportional to the radius $a$ of the fiber and depends in particular of the entrainment velocity $V$, the liquid surface tension $\gamma$ and the fluid viscosity $\eta$ through the so-called capillary number $\textrm{Ca}=\eta V/\gamma$, which compares the velocity to a fluid intrinsic velocity.
Thus, the thickness $h$ can be written as $h=a\,f(\textrm{Ca})$, where $f$ is a function which can be simplified under certain conditions on the capillary ranges.
Coating of complex fluids also received some attention, such as non-Newtonian fluids or water-surfactant solutions \cite{Quere1999,Rio2017}.

Due to the fiber curvature, the liquid film is unstable; a mechanism that was first characterized by Plateau \cite{Plateau1873} and rationalized by Lord Rayleigh in the 19th century \cite{Rayleigh1878}.
The liquid surface tension leads to the minimization of its surface area by breaking the film into a series of regularly spaced droplets.
The characteristic growing time of the Rayleigh-Plateau instability $\tau=12 \eta a^4/(\gamma h^3)$ depends significantly on the coating thickness.
For  non-Newtonian fluids, or for colloidal suspensions, the Rayleigh-Plateau instability can be delayed or suppressed
\cite{Boulogne2013b,Bauer2016,Boulogne2013a}.

Carroll \cite{Carroll1976} described the shape of axisymmetric drops on a fiber, which depends on the wetting properties of the material and the droplet volume with respect to the fiber size.
Based on Carroll's results, the drop height and the length of the wetted area can be used to characterize the wettability of the liquid on the fiber \cite{McHale1997,McHale1999,McHale2001,McHale2002}.
Beyond the axisymmetric shape, the equilibrium conformation of a drop on a fiber can also be a clam-shell: a drop sitting on a curved surface.
Experimental observations have evidenced that both barrel and clam-shell conformations can coexist and Chou {\em et al.} studied in detail the phase diagrams of droplet-on-fiber with or without gravity \cite{Chou2011}.

The aim of the paper is to describe the different evaporation regimes of the fiber coatings when the evaporation is dominated by the diffusion of water in the atmosphere.
Despite the fact that this is a model system with many industrial applications, there is a lack of theoretical rationalization on this subject.

In 1918, Langmuir \cite{Langmuir1918} explained for the first time that the total diffusive evaporation flux over a spherical droplet of size $R$ is not proportional to $R^2$, which means that, due to the curvature of the surface of the droplet, the evaporation flux is not proportional to the surface of the drop.
The case of a sessile drop evaporation (see~\cite{Cazabat2010, Brutin2015} for a review) is more complex because of the diverging evaporative flux at the triple line \cite{Lebedev1965,Picknett1977,Deegan1997} and several studies have been devoted to the nature of the substrate.
For instance, the case of a drop on a tilted surface  \cite{Timm2019}, on crossed fibers \cite{Boulogne2015a}, on superhydrophobic surfaces \cite{Stauber2015} or with complex wetting patterns \cite{Saenz2017} can be cited.
The situation of a droplet on a curved surface such as a convex or concave surface \cite{Paul2021,Shen2020} has been studied as well, the latter corresponding to a clam-shell on a fiber.

In this paper, we discuss the two limit cases of the diffusive evaporation of a liquid deposited on a fiber.
In a first part, we study the evaporation of a liquid cylinder deposited on a fiber that we call the {\em sleeve} configuration.
This configuration corresponds to a deposited layer observed at a time much smaller than the Rayleigh-Plateau characteristic time or when the instability is inhibited.
To do so, we will develop an analytical model to calculate the evaporation, valid for small and large aspect ratios.
These predictions are supplemented by a full 3D numerical simulation using finite element method.
In the second part, we study the evaporation of a droplet on a fiber which corresponds to destabilized coating on a fiber.
More precisely, we restricted the study to the axisymmetric barrel situation which is the common case, in particular for small contact angles of the liquid on the fiber.
We first numerically calculate the shape of a droplet around a fiber using Surface Evolver \cite{Brakke1992}, a surface minimization algorithm, whose results can be compared to the prediction by Carroll \cite{Carroll1976} and Chou {\em et al.} \cite{Chou2011} in absence of gravity.
By using finite element computations, we calculate the evaporation flux around the droplet, and show that except for very small liquid volume, the evaporation rate of the droplets is not significantly affected by the presence of the fiber, whatever the liquid contact angle is.

%%%%%%%%%%%%%%%%%%%%%%%
%
% SLEEVE
%
%%%%%%%%%%%%%%%%%%%%%%%

\section{Sleeve}\label{sec:sleeve}

\begin{figure}[htbp]
    \includegraphics[width=1\linewidth]{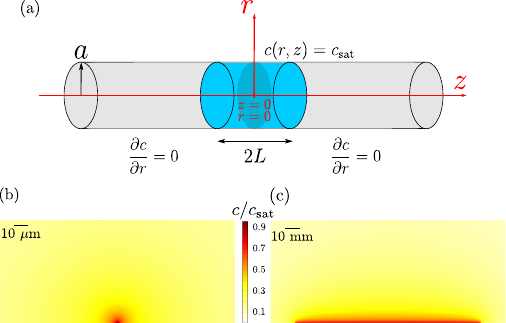}
    \centering
    \caption{(a) Notations used to describe a sleeve of length $2L$ on a fiber of radius $a$.
    The dimensionless length is defined as $\lambda = L/a$.
    Boundary conditions used to solve equation~\ref{eq:sleeve_laplace_equation} as well as the coordinate system are also represented schematically.
    Vapor concentration field for (b) $\lambda = 5 \cdot 10^{-3}$ and (c) $\lambda = 500$,  obtained from COMSOL for sleeves of radius $a= 125$~µm.
    }
    \label{fig:sleeve_img}
\end{figure}

We consider a sleeve of liquid of length $2L$ and of radius $a$ as depicted in figure~\ref{fig:sleeve_img}(a), such that the liquid cylinder has the same radius as the fiber.
A natural geometrical parameter is $\lambda = L / a$.

We assume that the liquid evaporation occurs in a diffusion-limited process in the stationary regime.
Thus, the concentration field $c$ in the gas phase is the solution of the Laplace equation $\triangle c = 0$, which reads in cylindrical coordinates

\begin{equation}
    \frac{1}{r}\frac{\partial}{\partial r}\left(r \frac{\partial c}{\partial r} \right) + \frac{\partial^2 c}{\partial z^2} = 0.
    \label{eq:sleeve_laplace_equation}
\end{equation}
In this geometry, the boundary conditions are (a) a saturated vapor concentration $c_{\rm sat}$ in the vicinity of the interface, \textit{i.e.} $c(r= a,z) = c_{\rm sat}$ for $|z| < L$, (b) the absence of evaporative flux at the surface of the fiber, \textit{i.e.} $\left. \frac{\partial c}{\partial r}\right\vert_{r=a} = 0 $ for $|z| > L$, and (c) a constant concentration far from the liquid, $\lim\limits_{r \to \infty}c = c_{\infty}$.

We provide in Supplementary Materials the derivation of analytical solutions to equation \ref{eq:sleeve_laplace_equation} in the limits of small and large aspect ratios $\lambda$.
The difficulty associated with this linear boundary-value problem (BVP) resides in the discontinuity of the boundary conditions (BC) at the inner boundary $r = a$.
More precisely, two different types of BCs, namely a Dirichlet-type BC and a Neumann-type BC are applied on disjoint complementary subdomains of the cylindrical surface located at $r = a$.
This difficulty prevents the use of the classical Fourier-Hankel analysis of the problem and, instead, the BVP is reduced to a set of dual integral equations.
Following the theory of Sneddon \cite{sneddon1995}, these coupled integral equations are then reduced to a single integral equation with a weak, logarithmic singularity that is solved analytically in the asymptotic cases $\lambda \ll 1$ and $\lambda \gg 1$.
From the resulting concentration field $c(r,z)$, we can compute the local evaporative flux $j(z)$ defined as $j(z) = - {\cal D} \left.\frac{\partial c}{\partial r}\right\vert_{r=a}$, where $\cal D$ is the diffusion coefficient of the vapor in the gas phase.

We also propose to solve equation~\ref{eq:sleeve_laplace_equation}  by using finite element method implemented in the proprietary software COMSOL multiphysics using Transport of Dilute Species physics in axisymmetric 2D geometry.
To compute accurately $j(z)$ while keeping the computational time reasonable, we divide the atmosphere in three concentric domains centered on the liquid.
The maximum mesh size in these domains is chosen according to the variations of concentration: the area near the three-phase contact is meshed with more refinement to capture properly the divergence of the concentration gradient at the contact line, and as the distance to the fiber increases, the maximum mesh size allowance is increased.
The size of the box describing the atmosphere is at least one hundred times larger than the largest of the lengths of the system in order to be considered as infinite with respect to the above boundary conditions.
With this approach, the computational time is reduced to few hours for the largest aspect ratios.
The numerical computations are performed for a vapor saturating concentration $c_{\rm sat} = 1.8\cdot10^{-2}$~kg/m$^{3}$, a vapor concentration far from the liquid $ c_\infty = 0$ and a diffusion coefficient of vapor in air ${\cal D} = 2.36\cdot10^{-5}$~m$^2$/s~\cite{Rankin2009}.
These values are chosen to be those of water evaporating at 20~$^\circ$C and zero relative humidity $\mathcal{R}_{\rm H} = c_\infty / c_{\rm sat}$.
The evaporative flux being proportional to ${\cal D} c_{\rm sat} (1-{\cal R}_{\rm H})$, these results can be easily generalized for any value of ${\cal R}_{\rm H}$, ${\cal D}$ and $c_{\rm sat}$.
Sleeves of different aspect ratios are generated by changing independently the length and the radius.

\subsection{Results}

In the limit $\lambda \ll 1$, we obtained analytically the local evaporative flux
\begin{equation} \label{eq:sleeve_j_small}
    %j_{\rm small}(z) = \frac{ j_0^{\rm small} }{ \sqrt{1-  \frac{z^2}{L^2}}}
    j_{\rm small}(z) =  j_0^{\rm small}  \left( 1-  \frac{z^2}{L^2} \right)^{-1/2},
\end{equation}
where $j_0^{\rm small}$ is the local flux at $z=0$ defined as
\begin{equation}\label{eq:sleeve_j0_small}
    j_0^{\rm small} = \frac{{\cal D} (c_{\rm sat} - c_{\infty})}{L(1- 2 \gamma_e - \ln\left(\frac{\lambda}{4}\right))},
\end{equation}
with $\gamma_e \approx 0.577$, the Euler gamma constant.

\begin{figure}[]\centering
    \includegraphics[width=\linewidth]{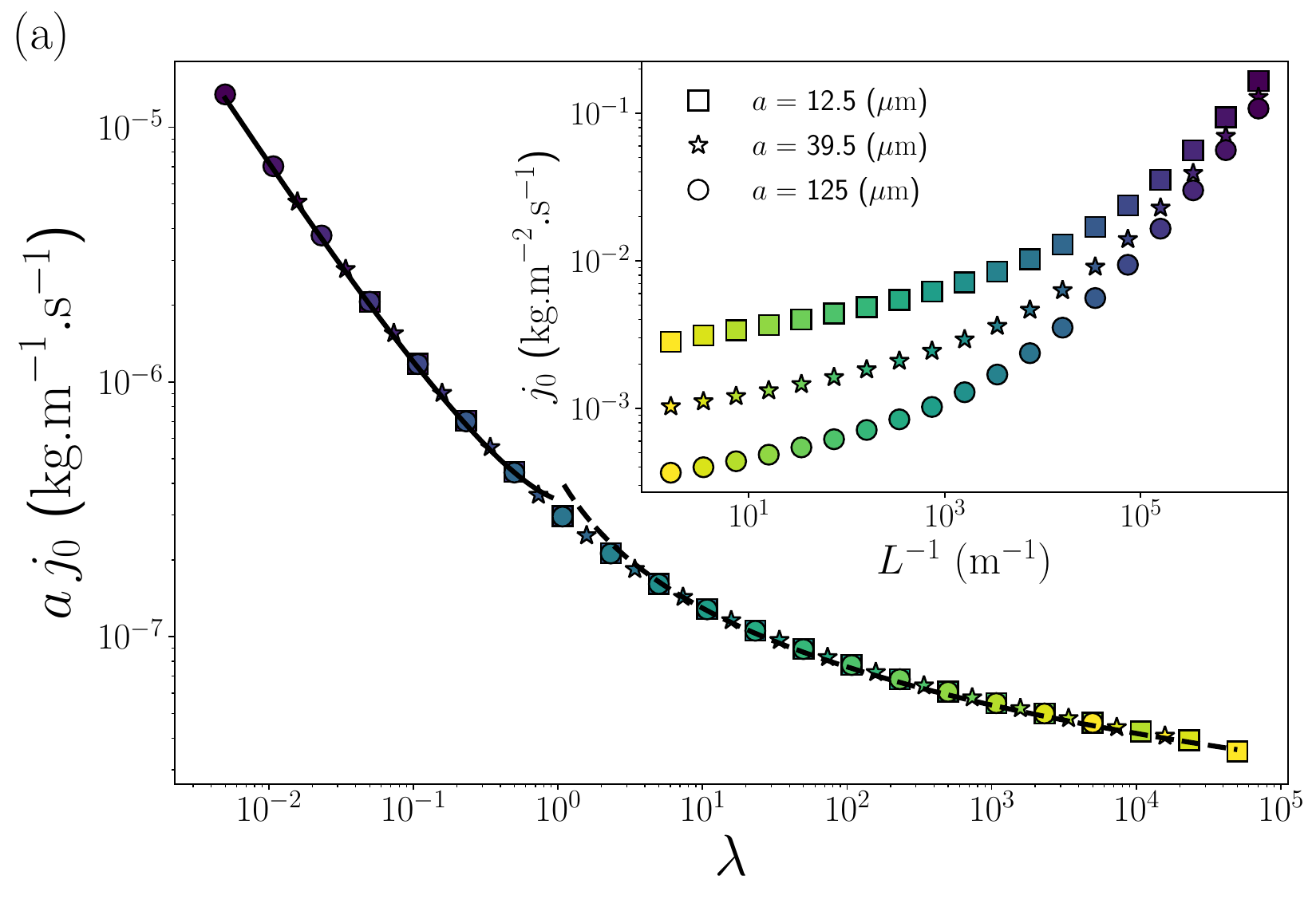}
    \includegraphics[width=\linewidth]{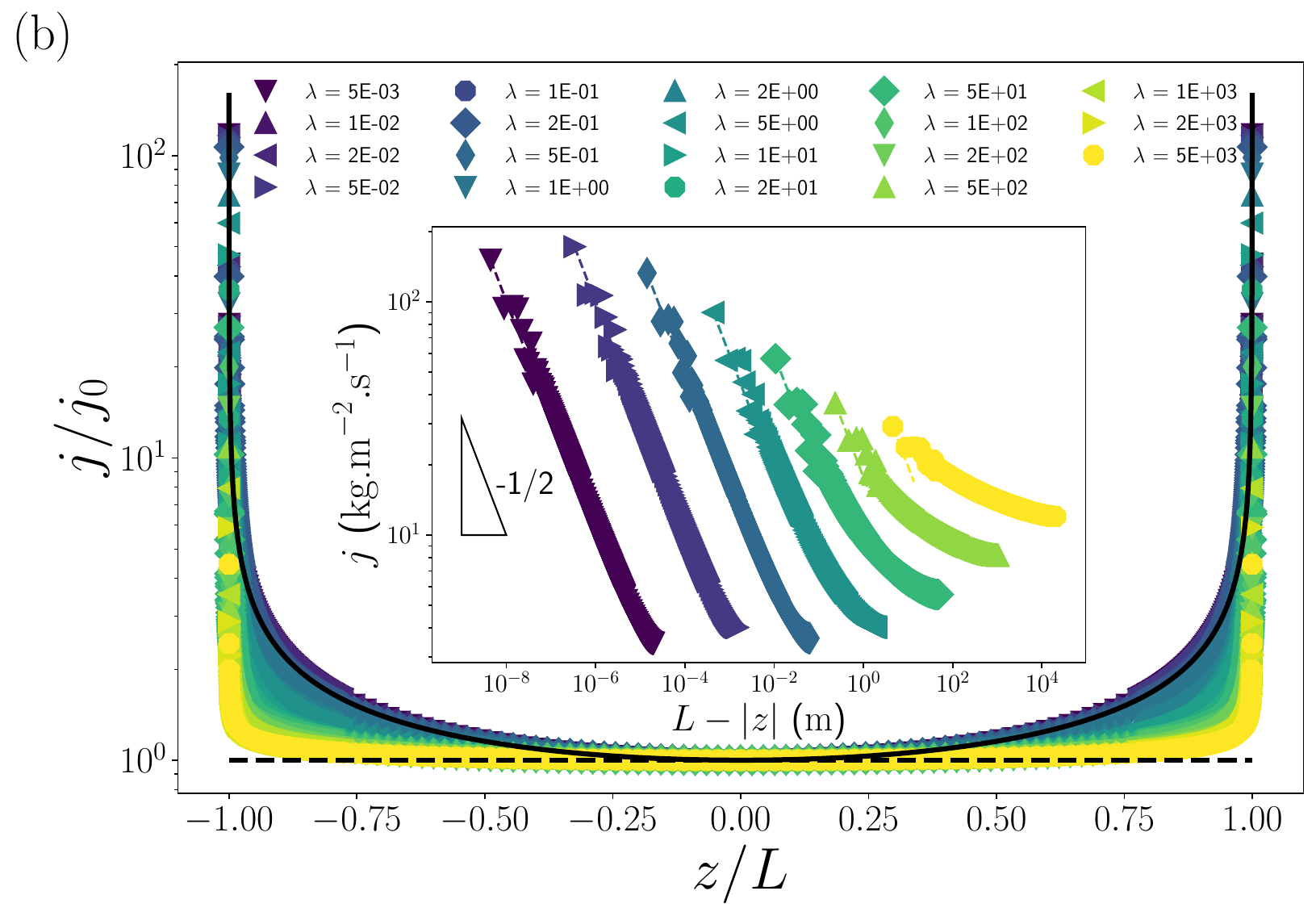}
    \caption{(a) Local evaporative flux the center of the sleeve $aj_0$ as a function of the dimensionless sleeve length $\lambda = L/a$ obtained from numerical computations for various sleeves lengths $L$ and three different radii $a$ (see caption).
    The black solid line corresponds to equation~\ref{eq:sleeve_j0_small} and the black dashed line corresponds to equation~\ref{eq:sleeve_j0_large} both multiplied by $a$.
    In the inset, the local flux at the center of the sleeve $j_0$ is plotted as a function $L^{-1}$ for the three sleeves radii mentioned above. (b) Dimensionless flux density $j/j_0$ as a function of dimensionless position $z/L$ along the sleeve.
    The points are obtained from numerical computation for $a = 125$~µm and various sleeves lengths.
    The black solid line corresponds to equation (\ref{eq:sleeve_j_small}) nondimensionalized by equation~\ref{eq:sleeve_j0_small}.
    The black dashed line corresponds to $j = j_0$ as predicted for infinitely long sleeve (Eq.~\ref{eq:sleeve_j0_large}).
    In the inset, the flux $j$ is plotted as a function of the distance to the contact line for $a = 125$~µm and various sleeves lengths.
    The curves are arbitrarily shifted for clarity.
    The color gradient represents the length $L$ of the sleeve on both graphics. }
    \label{fig:sleeve_j}
\end{figure}

Because the geometry is not reduced to a single lengthscale, equation \ref{eq:sleeve_j0_small} indicates that the flux depends both on the sleeve length $L$ and the aspect ratio $\lambda$, which is found numerically as shown in the Figure~\ref{fig:sleeve_j}a.
We find a good agreement between the numerical results and equation~\ref{eq:sleeve_j0_small}.

Equation \ref{eq:sleeve_j_small} indicates that the flux diverges at the edge between the liquid and the solid.
This behavior is confirmed numerically as observed in the Figure~\ref{fig:sleeve_j}b where numerical results obtained for various $\lambda$ are compared to the analytical prediction of equation~\ref{eq:sleeve_j_small} represented in solid black line.
This divergence is also obtained for sessile droplets, which indeed exhibit the same diverging expression in the limit $\theta \to 0$ \cite{Deegan1997}.

We also solve analytically equation~\ref{eq:sleeve_laplace_equation} for infinitely long sleeves \textit{i.e.} $\lambda \to \infty$.
Under this hypothesis, the system is invariant by translation along the fiber axis such that the local flux is uniform.
We can write
\begin{equation}
    j^{\rm large}_0 = \frac{{\cal D}(c_{\rm sat} - c_{\infty})\pi}{2a\left( 2- 2 \gamma_e + \ln{2} + \frac{\pi}{2} \ln{\lambda } \right)}. \label{eq:sleeve_j0_large}
\end{equation}
We extend this result to large but finite aspect ratios by arguing that the local evaporative flux remains mostly uniform, such that $ j_{\rm large}(z) \simeq j^{\rm large}_0$.

Again, we observe a good agreement between numerical and analytical results for the local flux at the center of the sleeve $j_0$ as shown in Fig.~\ref{fig:sleeve_j}a.
As for small $\lambda$, because of the cylindrical geometry, numerical results (Fig.~\ref{fig:sleeve_j}a) and equation~\ref{eq:sleeve_j0_large} indicate that the flux depends on both the sleeve radius $a$ and the aspect ratio $\lambda$.

For $\lambda > 1$, Figure~\ref{fig:sleeve_j}b shows that the increase of $\lambda$ leads to a localization of the divergence at the contact line.
For the largest tested aspect ratios, the flux per unit surface is nearly constant along the sleeve except close to the contact line.
However, as shown in the inset of Figure~\ref{fig:sleeve_j}b,  we still find close to the contact line the classical minus one-half power-law divergence even for large aspect ratio.
Indeed, mathematically, this divergence with a power $-1/2$ is the only one that is solution of equation~\ref{eq:sleeve_laplace_equation} to describe the divergence of the local flux at the contact line.
For $\lambda = 5\cdot 10^3$, numerical uncertainties prevent us from concluding on the expression of the divergence, but we expect that the classical square root law holds.

The total evaporative flux is defined as $Q = \int j(z)\,{\rm d}S $ where the integral is taken over the liquid-vapor surface area.
In the limit of small aspect ratios, the total flux writes
\begin{equation}\label{eq:sleeve_Q_small}
    Q_{\rm small} = \frac{2 {\cal D} (c_{\rm sat} - c_{\infty}) a \pi^2}{1- 2 \gamma_e - \ln\left(\frac{\lambda}{4}\right)}.
\end{equation}

\begin{figure}[]\centering
    \includegraphics[width=\linewidth]{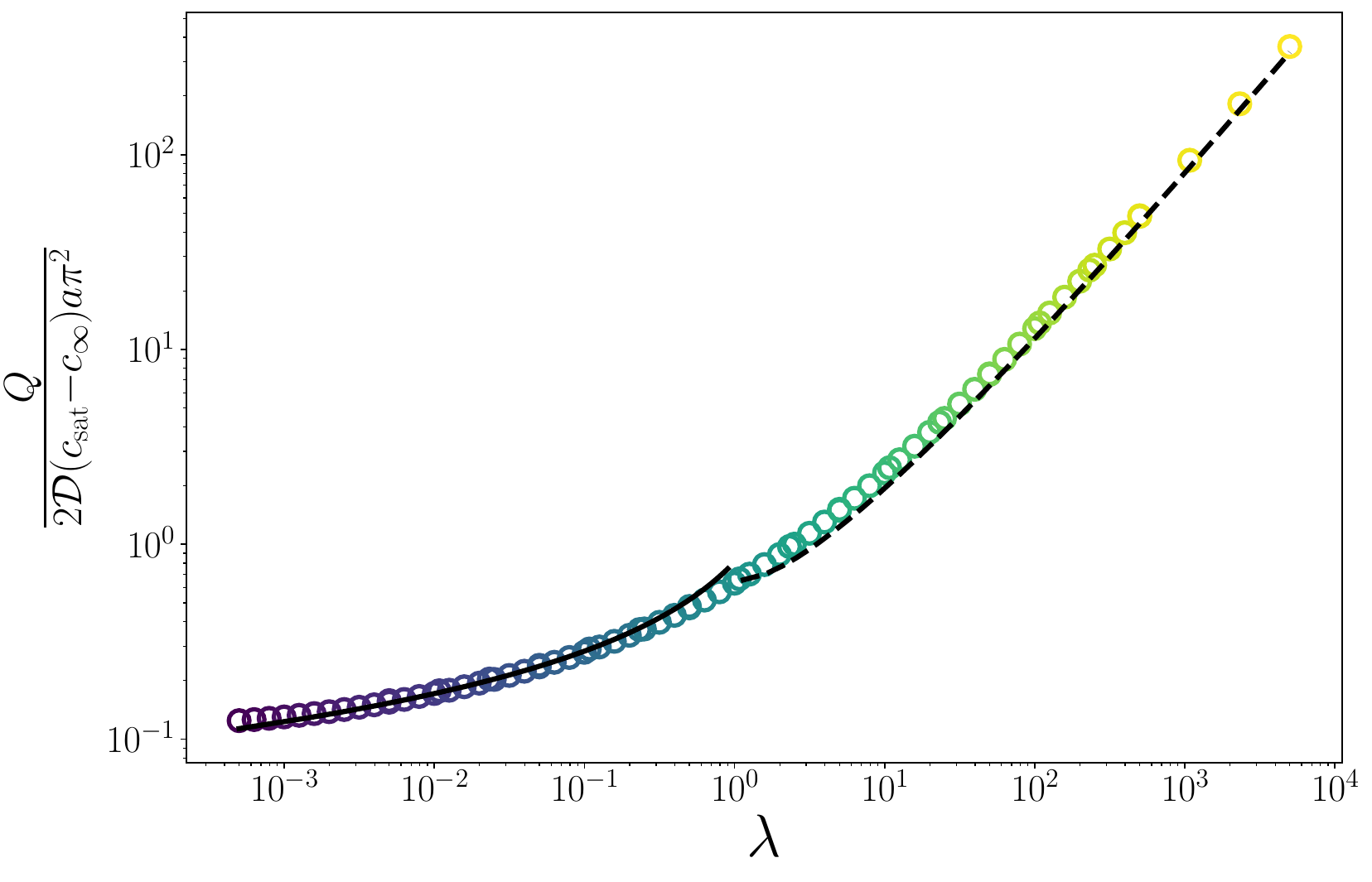}
    \caption{ Total dimensionless flux as a function of the dimensionless sleeve length $\lambda = L/a$ obtained from numerical computations for $a = 125$~µm. The solid black line corresponds to equation~\ref{eq:sleeve_Q_small} and the dashed black line to equation~\ref{eq:sleeve_Q_large}.}
    \label{fig:sleeve_Q_vs_lambda}
\end{figure}

In the limit of large aspect ratio, we neglect the contribution of the divergence.
Indeed, this divergence is localized and contributes weakly once integrated to the total flux \cite{Boulogne2017a}.
Thus, the total evaporative flux writes $Q_{\rm large} = j^{\rm large}_0 \, 4 \pi a L$, which gives
\begin{equation}\label{eq:sleeve_Q_large}
    Q_{\rm large} = \frac{2 {\cal D} (c_{\rm sat} - c_{\infty}) \lambda a \pi^2}{2- 2 \gamma_e + \ln{2} + \frac{\pi}{2} \ln{\lambda}}.
\end{equation}
As shown in figure~\ref{fig:sleeve_Q_vs_lambda}, where we plot the dimensionless total flux $Q/(2 {\cal D} (c_{\rm sat} - c_{\infty}) a \pi^2 )$ as a function of the aspect ratio $\lambda$, we observe an excellent agreement between the numerical results and the analytical model.
The good agreement between numerical results and theory at large aspect ratio demonstrates the small effect of the edge contribution to the local flux, which is due to the localization of the divergence when $\lambda$ increases.
Indeed, the more localized the divergence, the more constant the local flux $j$ can be considered and the less effect the variation of $j$ with $z$ has on the total integrated flux $Q$.

As a final note, the model described here is valid asymptotically for $\lambda \ll 1$ and $\lambda \to \infty$ but still provides an excellent description of both the local flux in the center of the liquid $j_0$ (Fig.~\ref{fig:sleeve_j}a) and the total evaporative flux $Q$ (Fig.~\ref{fig:sleeve_Q_vs_lambda}) for $\lambda \sim 1$.

In the next section, we present the method to perform the numerical computation of equation \ref{eq:sleeve_laplace_equation} with the associated boundary conditions for axisymmetric droplets on fibers.
We then discuss these results and compare them to those obtained for sleeves and spherical drops.

%%%%%%%%%%%%%%%%%%%%%%%
%
% DROP on FIBER
%
%%%%%%%%%%%%%%%%%%%%%%%

\section{Axisymmetric drop on a fiber}

Due to the curvature of the substrate, a droplet of perfectly wetting liquid put on a fiber does not necessarily adopt a sleeve morphology and a macroscopic axisymmetric drop with a vanishing contact angle can exist~\cite{Carroll1976,McHale1997,McHale1999,McHale2001,McHale2002,Chou2011}.
This morphology is called a barrel-shaped droplet and is stable for low contact angle $\theta$ and quite large volume with respect to $a^3$.
For small volume and/or large contact angle a droplet on a fiber adopt a clam-shell morphology.
Here, we choose to focus on barrel shaped droplets.
To study the evaporation of such droplets, we vary independently the drop volume $\Omega$ at constant fiber radius, $a = 125$~µm, and the contact angle $\theta$.
In contrast to the sessile droplet for which the system is completely defined by the choice of only two parameters, $\Omega$, and $\theta$, the equilibrium profile of a drop on a fiber depends on $\Omega$, $\theta$ and $a$.
Once two dimensionless parameters, say $\theta$ and $\Omega/a^3$ are chosen, the liquid adopts its equilibrium shape as represented in figure~\ref{fig:drop_profiles_and_notations}a, where the profile of the drop is defined by $h(z)$.
We denote the height of the liquid at the apex $h_0 = h(z = 0)$ and the wetted length $2L$.
We define two dimensionless parameters, the dimensionless length $\lambda = L/a$ and the drop aspect ratio $\mathcal{H} = h_0/L$.

The barrel shape morphology exists for a limited  dimensionless parameter space ($\lambda,\mathcal{H}$), which restricts the studied range of these parameters.
Due to the complexity of the problem arising from the drop shape, this part of the study is performed numerically.

In the next section, we present the method to get numerically the drop equilibrium profile and then to solve equation~\ref{eq:sleeve_laplace_equation} for this system.

\subsection{Numerical procedure}

\begin{figure}[]\centering
    \includegraphics[width=\linewidth]{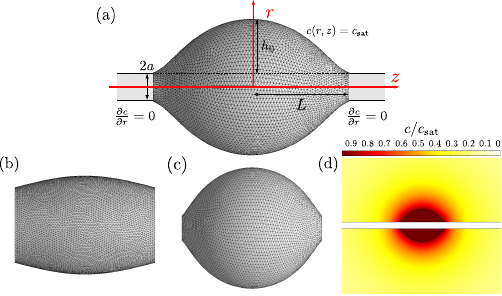}

    \caption{Surface Evolver profiles of axisymmetric drops on fiber of radius $a  = 125$~µm.
    (a) Drop of volume $\Omega = 1$~µL and contact angle $\theta = 10^\circ$.
    The wetted length is $L$ and the height of the drop $h_0$ is defined between the surface of the fiber and the apex of the drop.
    (b) Drop of volume $\Omega = 0.01$~µL and contact angle $\theta = 10^\circ$.
    (c) Drop of volume $\Omega = 1$~µL and contact angle $\theta = 45^\circ$.
    (d) Vapor concentration field from COMSOL for profile given in (a) $a = 125$~µm, $\Omega = 1$~µL, $\theta = 10^\circ$.
    }
    \label{fig:drop_profiles_and_notations}
\end{figure}

In contrast to the numerical resolution of the sleeve configuration, our numerical procedure for the drop on fibers is decomposed in two steps.
First, we used Surface Evolver \cite{Brakke1992}, a computer program that minimizes the energy of a surface subject to constraints, to obtain the meshed surface of the drop on a fiber.
The parameters are the dimensionless drop volume $\Omega/a^3$  and the liquid contact angle $\theta$.
The convergence of the minimization process is testified by comparing the drop height $h_0$ and length $L$ (Fig.~\ref{fig:drop_profiles_and_notations}) with the analytical predictions made by Carroll \cite{Carroll1976}.
Illustrations of the resulting shapes are presented in figure~\ref{fig:drop_profiles_and_notations}(a-c).
The second step consists in computing the vapor concentration field with COMSOL multiphysics in a similar manner as for the sleeves except that the simulations are performed in 3D in order to easily import the profiles obtained with Surface Evolver.
With COMSOL, the Surface Evolver meshes are converted in 2D to keep only the boundary surface.
A 3D drop is then reconstructed to ensure the compatibility with other elements of the geometry and to mesh properly the contact between the drop and the fiber.
The boundary conditions are the same as for the sleeve, the vapor concentration is equal to the saturating vapor concentration near the interface and is constant far from the droplet \textit{i.e.} $c(r = h, z) = c_{\rm sat}$, $\lim\limits_{r \to \infty}c = c_{\infty}$.
Initial conditions are $c=c_\infty$ for the concentration in the atmosphere  and $c=c_{\rm sat}$ inside the drop.
There is no vapor flux normal to the surface of the fiber $\left. \frac{\partial c}{\partial r}\right\vert_{r=a} = 0 $ for $|z| > L$ (Fig.~\ref{fig:drop_profiles_and_notations}a).
As for the sleeve case, the dimension of the box describing the atmosphere is taken to be at least one hundred times larger than the longest lengths of the system to ensure $c = c_\infty$ far from the droplet.

\begin{figure}[]\centering
    \includegraphics[width=\linewidth]{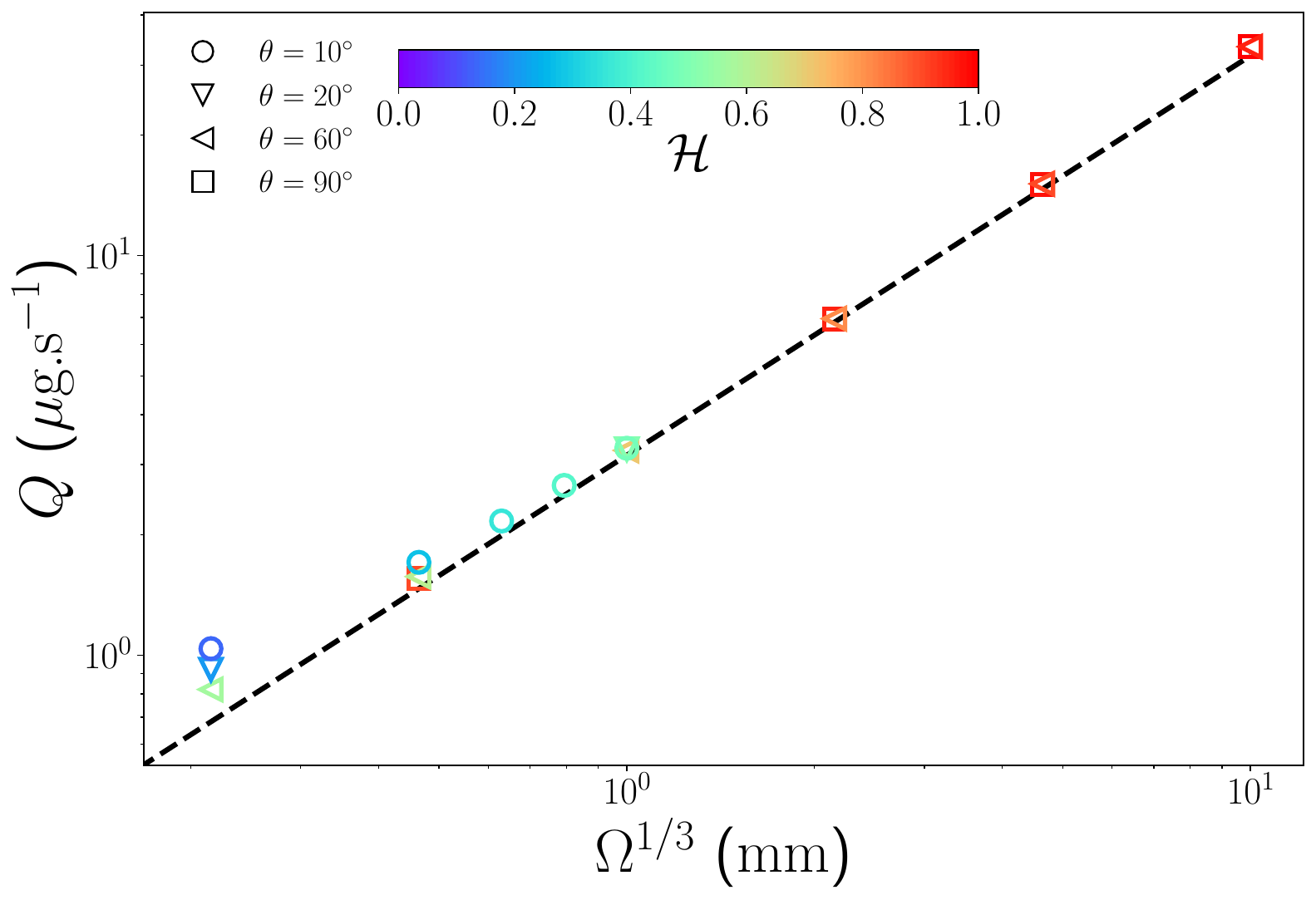}
    \includegraphics[width=\linewidth]{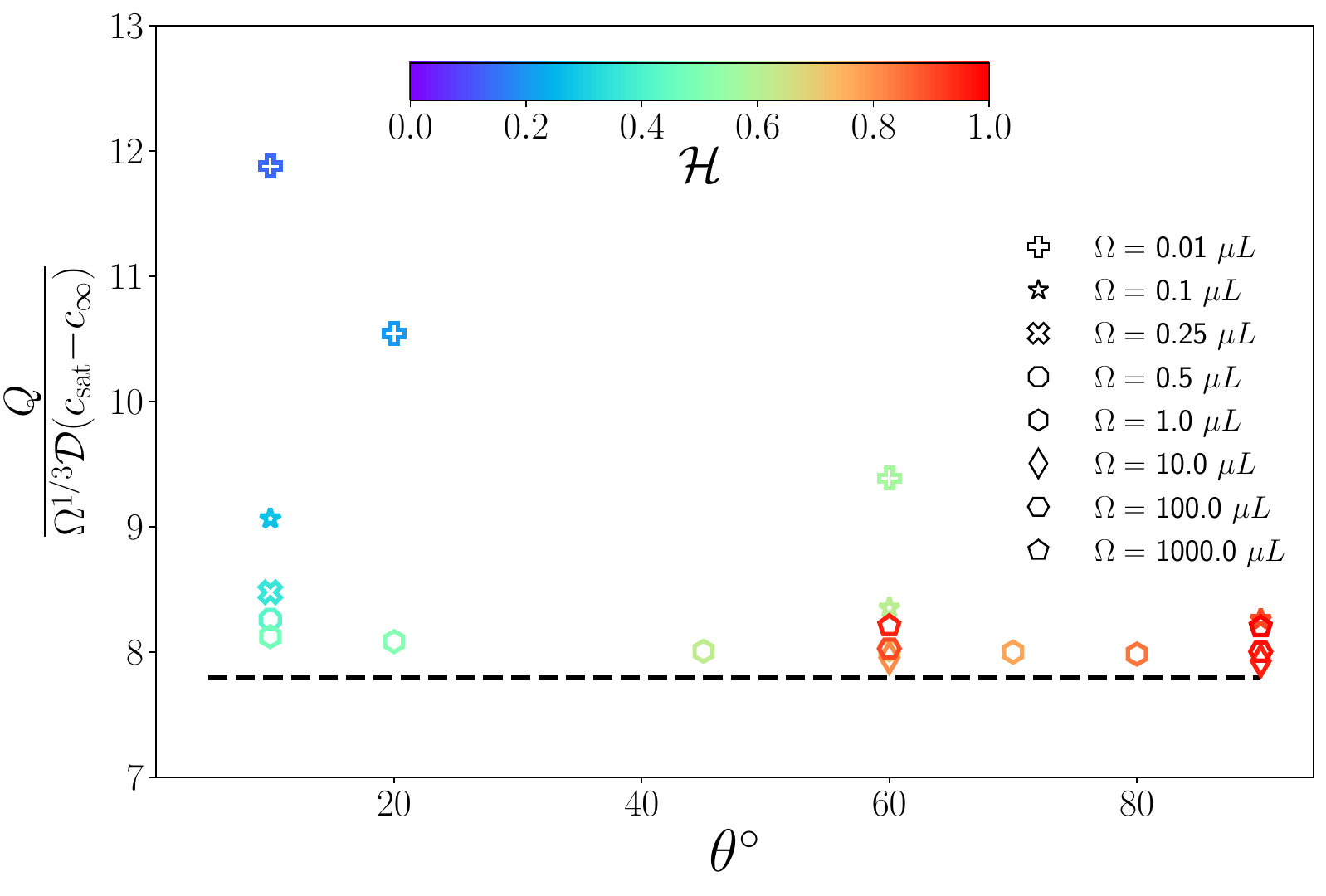}
    \caption{(a) Total evaporative flux of droplets on fiber as a function of $\Omega ^{1/3}$. (b) Total dimensionless evaporative flux as a function of contact angle.
    The  points are numerical results for evaporating drops on fiber for different contact angles. Colors indicate the variation of the droplet aspect ratio ${\cal H} = h_0/L$.
    The dashed black line is the evaporative flux of a sphere (Eq.~\ref{eq:Q_sphere}) in the same condition as the numerical simulations.}
    \label{fig:drop_Q_vs_omega_1_tiers}
\end{figure}

\begin{figure}[]\centering
    \includegraphics[width=\linewidth]{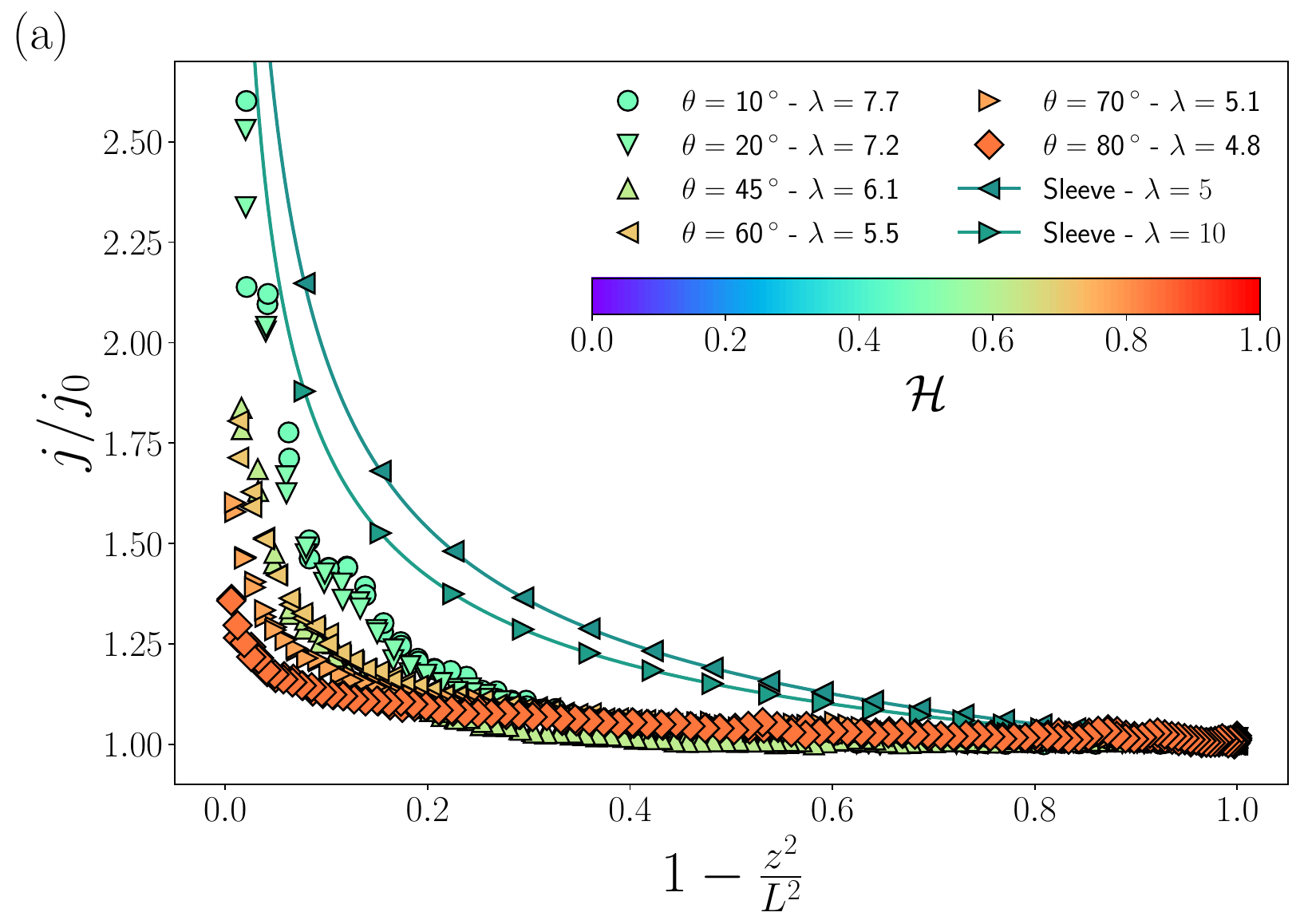}
    \includegraphics[width=\linewidth]{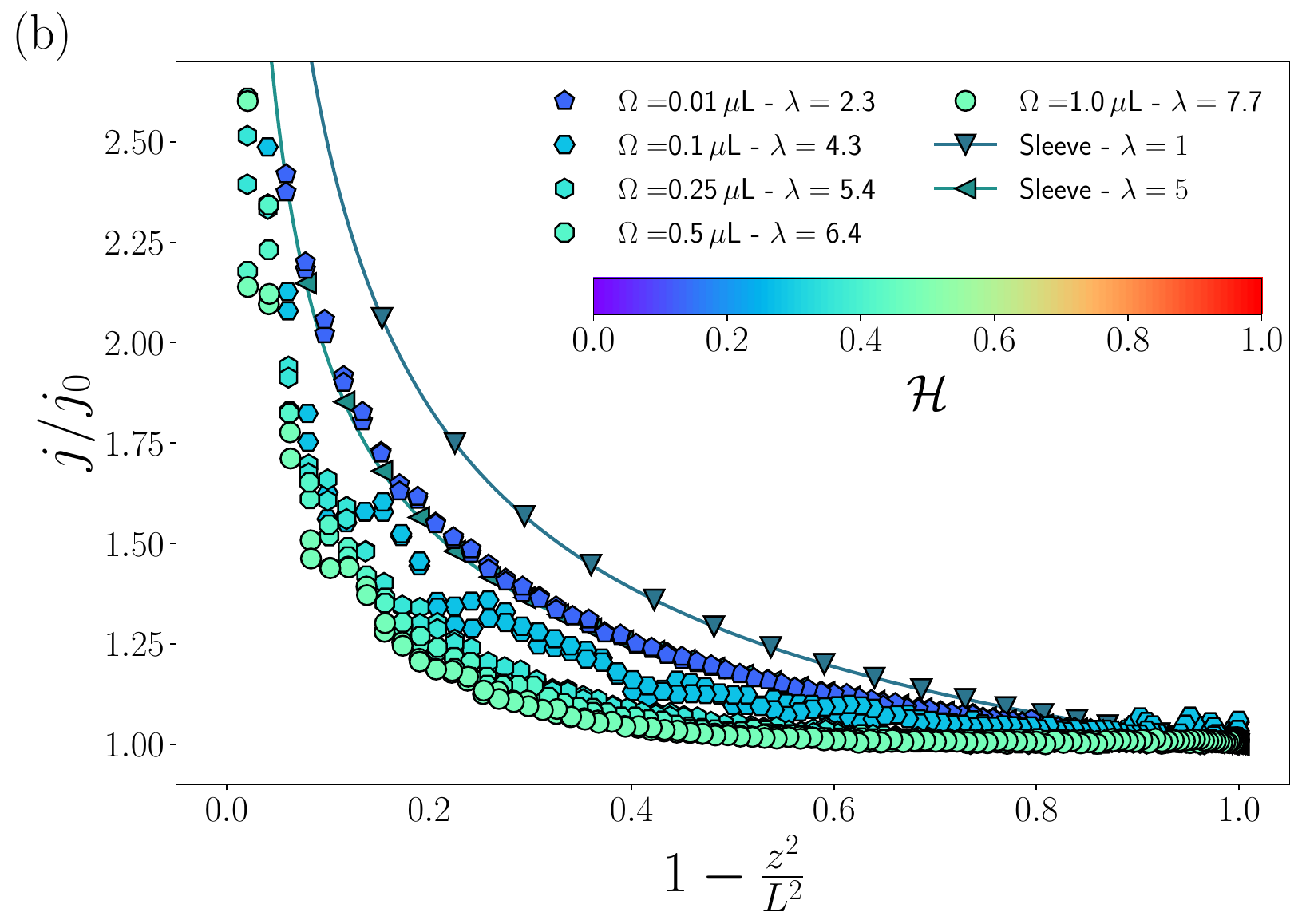}
    \caption{Dimensionless evaporative flux per unit surface $j/j_0$ as a function of $1 - (z/L)^2$.
    The points are obtained by numerical simulations for droplets of volume $\Omega$ and contact angle $\theta$ placed on a fiber of radius $a = 125$~µm.
    The dimensionless wetted length $\lambda = L/a$ of the droplets are also given and the results for sleeves with the same $\lambda$ are plotted for comparison.
    The colors represent the droplet aspect ratio ${\cal H} = h_0 / L$.
    (a) Results obtained for $\Omega = 1$~µL and contact angle varying from $10$ to $80^\circ$ corresponding to $\lambda$ ranging from $5$ to $8$ and aspect ratio ${\cal H}$ between 0.5 and 0.8.
    (b) Results obtained for $\theta = 10^\circ$ and droplets volumes varying from $0.01$~µL to $1$~µL corresponding to $\lambda$ ranging from $2$ to $8$ and aspect ratio ${\cal H}$ varying from $0.1$ for the smallest volume tested to $0.5$.}
    \label{fig:drop_j_vs_z}
\end{figure}

\subsection{Results}
First, we want to understand how the presence of the fiber influences the evaporation speed.
To do so, we obtained numerically the total evaporative flux $Q$ of barrel-droplets on fibers of radius $a = 125$~µm.
Numerical simulations are performed for various drop volumes $\Omega$ and contact angles $\theta$.
We recall that we limit ourselves to cases of barrel-shaped droplets, which only exists for drops of sufficiently large volumes that are placed on hydrophilic substrates~\cite{Chou2011}.

The results are shown in figure~\ref{fig:drop_Q_vs_omega_1_tiers}a where we plot $Q$ as a function of $\Omega^{1/3}$ and in figure~\ref{fig:drop_Q_vs_omega_1_tiers}b representing the dimensionless evaporative flux as a function of the contact angle.
The total evaporative flux of a spherical drop~\cite{Langmuir1918},
\begin{equation}\label{eq:Q_sphere}
    Q_{\rm sphere} \: = \: (48 \pi^2)^{1/3} \ {\cal D} \ (c_{\rm sat} - c_\infty) \ \Omega^{1/3},
\end{equation}
is also represented in dashed black line for comparison.
We observe that the evaporation rate of a barrel-shaped drop on a fiber is similar to that of a sphere for all the contact angles studied except for the smallest volumes tested. This means that varying the contact angle has a negligible effect on the evaporative flux except for small volume drops. We also note that, for small volumes, the deviation from the sphere is increasing as the contact angle is getting smaller, the largest variation being observed for the 0.01~µL drop having a contact angle of $10^\circ$ whose profile is given in figure~\ref{fig:drop_profiles_and_notations}b.
From Fig.~\ref{fig:drop_Q_vs_omega_1_tiers}b, we can quantify the deviation from the sphere, which varies from a few percent for $\Omega \geq 0.1µL$ to around 35$\%$ for $\theta = 10^\circ,~ \Omega = 0.01$~µL

We can also note that decreasing contact angle has a small effect on the drop aspect ratio $\mathcal{H}$ except for small volume drops.
This is due to the curvature of the substrate that allows the existence of a macroscopic drop ($\mathcal{H} \sim 1$) even for small contact angles.

To understand these observations in more details, we consider the local evaporative flux of barrel-shaped droplets.
First, we focus on the effect of the contact angle.
In figure~\ref{fig:drop_j_vs_z}a, we plot the local flux $j$ as a function of the position along the interface $z$ for a droplets of volume 1~µL with various contact angle corresponding to aspect ratios ${\cal H} \simeq 1$.
This figure shows that the flux $j$ diverges in close vicinity of the drop contact line.
The decrease of the contact angle has a very small effect on the local evaporative flux and this effect is significant only near the contact line.
We also compare the results obtained for barrel drops for which $\lambda \in [5 ; 10]$ with those of the sleeves having the same dimensionless length.
These sleeves have an aspect ratio $\lambda > 1$, which means that the divergence of the local flux has already the localization effect described in the previous Section\ref{sec:sleeve}.
The comparison between the drops and the sleeves highlight the effect of the drop profile curvatures.
The longitudinal curvature of the surface is estimated by $\mathcal{H}$, with $\mathcal{H} =1 $ corresponding to a spherical drop whereas $\mathcal{H} = 0$ is a sleeve.
Fig.~\ref{fig:drop_j_vs_z}a demonstrates that, due to the curvature of the drop surface, the divergence of the evaporation flux of a drop on a fiber is even more localized at the triple line than for the sleeve.
We propose a phenomenological equation for the local flux $j(z)$ of drops on fibers defined as
\begin{equation}
    j(z) = j_0\left[\beta\left( 1 - \frac{z^2}{L^2}  \right)^{-\alpha} + (1 -  \beta)\right],
\end{equation}
where $\alpha$ and $\beta$ are positive adjustable parameters.
Figure~4 provided in the Supplementary Materials shows fitted data for 1~µL drops for different contact angle.
The good agreement between numerical results and fitted curves shows that, unlike sessile drops, the local evaporative flux of a drop on a fiber cannot be described by a simple power law.
If we compare the local flux of barrel-shaped droplets to the local flux of a spherical droplet (Fig~2a-b in Supplementary Materials) we see that the local flux at the center of the barrel-shaped droplets is about $1.5$ times smaller than the evaporative flux of a sphere.
Thus, barrel-shaped drops are evaporating at the same speed as spherical droplets because the difference in local fluxes at the center of the drop compensates the localized divergence of the local flux of barrel-shaped droplets.

Finally, to understand the difference observed between the total evaporative flux of a spherical drop and the one of a small volume wetting barrel-drop on a fiber, we plot in figure~\ref{fig:drop_j_vs_z} the  local evaporative flux as a function of the position along the interface for drops of small contact angle $\theta = 10^\circ$ and different volumes.
This difference is significant for small volumes and small contact angles, \textit{i.e.} when $\mathcal{H}$ vanishes.
In this case, the geometry is similar to a to liquid cylinder and we obtain results comparable to those expected for a sleeve.
Nevertheless, the drop, which has a dimensionless length $\lambda \approx 2.3$ can be well compared to the sleeve for a greater aspect ratio, $\lambda = 5$.
We interpret that although $h_0$ tends to 0, the liquid thickness still has a significant effect on the evaporative flux.

%%%%%%%%%%%%%%%%%%%%%%%
%
% CONCLUSION
%
%%%%%%%%%%%%%%%%%%%%%%%
\section{Conclusion}

In this paper, we studied the curvature effect on the evaporation rate of a liquid deposited on a fiber when the evaporation is isothermal and purely diffusive.
Two particular situations where analyzed: {\em a sleeve} of liquid of size $L$ deposited on fiber of radius $a$ and a droplet of volume $\Omega$ and wetting contact angle $\theta$ on the same fiber.
The sleeve is obtained when a liquid fiber is coated by a liquid layer before the onset of destabilization of the Rayleigh-Plateau instability.
The droplet, and more particularly the {\em axisymmetric barrel shape} that we studied here, is encountered  in the late stage of destabilization when the liquid coating has been destabilized in a series of liquid pearls.

For the evaporation of the sleeve we obtained an analytical calculation of the local evaporation rate along the sleeves and the full evaporation rate in the case of small and large aspect ratios.
For large aspect ratios, the evaporation rate is almost uniform along the sleeve except near the edges, where we recover the same power divergence than the one observed for sessile droplets at low contact angles.
For small aspect ratio, the role of the edges progressively becomes more and more important, the evaporation rate is varying significantly along the sleeve since the power law divergence invades all the sleeve.
We compared our analytical calculation to finite element computations and shown that our asymptotic calculations captures quantitatively the simulations even in the regime where the aspect ratio is close to one.

For the evaporation of the liquid axisymmetric barrels, we performed numerical simulations in order to calculate precisely the effect of the fiber on the evaporation rate.
We have evidenced that, for drops of volume $\Omega \geq 0.1$~µL corresponding to dimensionless volume $\Omega/a^3 \geq 50$, the evaporation rate is almost independent of the wetting contact angle and that the droplet evaporates as a sphere of the same volume.
More precisely, the evaporation flux diverges near the triple line, but due to the barrel shape, the divergence is strongly localized close to the edges and the evaporation rate is nearly constant.
Indeed the divergence close to the edge compensates fortunately the fact that the longitudinal curvature of the surface at the apex, estimated by $\cal H$, is not equal to the one of a spherical drop.

This study provides precise calculations that capture the drying dynamics of the two important morphologies on a fiber under the assumption of an isothermal evaporation process.
They both emphasize the localization of the evaporation divergence close to the triple line, in contrast to sessile drops.
To complete this analysis of the evaporation of a liquid coated on a fiber, it is necessary in a future work to focus more precisely on what happens after the destabilization of the sleeve into a series of regularly spaced drops.
Indeed, additional studies would be necessary to understand and quantify the mutual influence of the drops on each other.

\section*{Acknowledgments}
We thank Saint-Gobain and ANRT for funding this study and J. Delavoipière, M. Lamblet, and B. Sobac for useful discussions. We are also grateful to C. Dubois from COMSOL for providing support.

\bibliography{biblio}

\bibliographystyle{unsrt}

\newpage
\clearpage


\section*{Supplementary material}

# Supplementary materials: Evaporation of a liquid coated on a fiber

M. Corpart[1], J. Dervaux[2], C. Poulard[1], F. Restagno[1], and F. Boulogne[1]

[1]Université Paris-Saclay, CNRS, Laboratoire de Physique des Solides, 91405, Orsay, France.
[2]Université de Paris, Laboratoire Matière et Systèmes Complexes, UMR 7057 CNRS, F-75013 Paris, France

In these materials, we report the details of our calculations of the Laplace equation in Section 1. Then, we show additional data obtained numerically on the evaporative flux in Section 2.

## 1 Evaporation of liquid sleeves

In this Section, we solve the diffusion equation of an evaporating liquid sleeve with the boundary conditions specified in the main article that we recall briefly here.

### 1.1 Problem

All lengths are normalized by the radius $a$ of the fiber. The dimensionless total length of the sleeve is denoted $2\lambda$. In physical units, its total length is thus $2\lambda a$. We also introduce the rescaled concentration $\tilde{c} = (c - c_\infty)/(c_{\text{sat}} - c_\infty)$ which now has values in the interval $[0, 1]$. We thus solve the following boundary value problem:

$$\triangle \tilde{c} \equiv \frac{1}{r}\frac{\partial}{\partial r}\left(r\frac{\partial \tilde{c}}{\partial r}\right) + \frac{\partial^2 \tilde{c}}{\partial z^2} = 0 \tag{1}$$

$$\lim_{r \to \infty} \tilde{c} = 0 \tag{2}$$

$$\tilde{c}(r=1, z) = 1 \text{ for } z < \lambda \tag{3}$$

$$\left.\frac{\partial \tilde{c}}{\partial r}\right|_{r=1} = 0, \text{ for } z > \lambda. \tag{4}$$

The difficulty associated with this linear boundary value problem resides in the discontinuity of the boundary condition at $r = 1$. More precisely, two different types of boundary conditions, namely a Dirichlet-type boundary condition (3) and a Neumann-type boundary condition (4) are applied on disjoint complementary subdomains of the surface defined as $r = 1$. This difficulty prevents the use of the classical Fourier-Hankel analysis of the problem and a dual integral approach must be employed.

### 1.2 Solution - theoretical details

Let us introduce the following cosine transform $\bar{c}(k, r)$ of the dimensionless concentration field $\tilde{c}(r, z)$:

$$\tilde{c}(r, z) \equiv \int_0^\infty \bar{c}(k, r) \cos(kz) \mathrm{d}k. \tag{5}$$

Substituting the transformation (5) in the Laplace equation (1) results in an ordinary differential equation for $\bar{c}(k, r)$,

$$\frac{1}{r}\frac{\partial}{\partial r}\left(r\frac{\partial \bar{c}}{\partial r}\right) - k^2 \bar{c} = 0. \tag{6}$$

This equation can be solved by a standard separation of variable and has two independent Bessel-like solutions but only one of them vanishes in the limit $r \to \infty$ and thus satisfies (2). This solution can be written as

$$\bar{c}(k, r) = C(k) K_0(kr), \tag{7}$$

where $K_0$ is the modified Bessel function of the second kind of integer order 0. The remaining task is to find the function $C(k)$ such that (3) and (4) are satisfied. Plugging the transform (5) together with the solution (7) into the two boundary conditions (3) and (4) leads to the following system know as dual integral equations

$$\int_0^\infty C(k) K_0(k) \cos(kz) \mathrm{d}k = 1, \text{ for } z < \lambda, \tag{8}$$

$$\int_0^\infty k C(k) K_1(k) \cos(kz) \mathrm{d}k = 0, \text{ for } z > \lambda, \tag{9}$$

where $K_1$ is the modified Bessel function of the second kind of integer order 1. Introducing the function

$$\kappa(k) = \frac{K_0(k)}{K_1(k)}, \tag{10}$$

as well as the function $D(k) = k C(k) K_1(k)$, the dual integral equations (8, 9) can be recast in the canonical form:

$$\int_0^\infty \frac{\kappa(k)}{k} D(k) \cos(kz) \mathrm{d}k = 1 \text{ for } z < \lambda, \tag{11}$$

$$\int_0^\infty D(k) \cos(kz) \mathrm{d}k = 0, \text{ for } z > \lambda. \tag{12}$$

In order to solve (11, 12), we choose the singular integral equation approach and start by noticing that the second equation (12) is automatically satisfied if we define the auxiliary function $g(u)$ as:

$$D(k) = \int_0^\lambda g(u) \cos(ku) \mathrm{d}u \tag{13}$$

$$= \int_0^1 \lambda \bar{g}(u) \cos(\lambda k u) \mathrm{d}u, \tag{14}$$

where $\bar{g}(u) = g(\lambda u)$. The result above follows directly from Fourier's inversion theorem. Substituting (14) into (11) leads to a double integral equation, which reads, after inverting the order of integration and introducing the reduced variable $z' = \lambda z$:

$$\int_0^1 \bar{g}(u) \left( \int_0^\infty \frac{\kappa(k)}{k} \cos(\lambda k z') \cos(\lambda k u) \mathrm{d}k \right) \mathrm{d}u = \frac{1}{\lambda}, \text{ for } z' < 1. \tag{15}$$

Using usual trigonometric relations and the fact that $g$ is an even function, we may recast (15) as:

$$\int_{-1}^1 \bar{g}(u) \mathcal{L}(\lambda(z' - u)) = \frac{2}{\lambda} \text{ for } |z'| < 1, \tag{16}$$

where the kernel $\mathcal{L}(x)$ is given by the following integral:

$$\mathcal{L}(x) = \int_0^\infty \frac{\kappa(k)}{k} \cos(kx) \mathrm{d}k. \tag{17}$$

Unfortunately, the kernel (17) cannot be expressed in closed form. In order to make further progress, let us focus on the asymptotic behavior of equations (16)-(17).

## 1.3 Small sleeve $\lambda \ll 1$

In the limit where the length of the sleeve is much smaller than the radius of the sleeve, *i.e.* when $\lambda \ll 1$, we also have $\lambda(z' - u) \ll 1$ since $\{z', u\} \in [-1, 1]$. We may thus use an approximation of $\mathcal{L}(x)$ valid at small $x$. In order to obtain such an approximation, we first split the kernel (17) into two integrals:

$$\mathcal{L}(x) = \int_0^1 \frac{\kappa(k)}{k} \cos(kx) \mathrm{d}k + \int_1^\infty \frac{\kappa(k)}{k} \cos(kx) \mathrm{d}k \tag{18}$$

Next we replace $\kappa(k)$ by its small-$k$ approximation (resp. large-$k$ approximation) in the first (resp. second) integral above. Using the following approximations at small and large $k$:

$$\kappa(k) = -\gamma_e - \ln \frac{k}{2} + \mathcal{O}\left(k^2 \ln k^2\right) \text{ for } k \to 0$$

$$\kappa(k) = 1 + \mathcal{O}\left(\frac{1}{k}\right) \text{ for } k \to \infty$$

we obtain the following approximate expansion for the kernel $\mathcal{L}(x)$

$$\mathcal{L}(x) \approx -\ln |x| + \sum_{n=0}^{n=\infty} b_n |x|^n \text{ for } x \ll 1 \tag{19}$$

where the first few terms of the development are given in table 1.3 and $\gamma_e = 0.577...$ is the Euler gamma constant. Note that the value of the coefficient $b_n$'s depends on the underlying approximation for $\kappa$. We now look for a series solution of (16) in the following form:

$$\bar{g}(x) = \frac{1}{\lambda} \sum_{n=0}^\infty \lambda^n \bar{g}_n(x, \ln \lambda) \tag{20}$$

Inserting the previous relations (19) and (20) into the singular integral equation, we obtain the following recurrent chain of integral equation defined in the domain $|z'| < 1$:

$$\int_{-1}^1 \bar{g}_n(u, \ln \lambda) \left(\ln |z' - u| - b_0 + \ln \lambda\right) \mathrm{d}u = f_n(z', \ln \lambda), \tag{21}$$

where the right hand side of (21) is given by:

$$f_0(z', \ln \lambda) = -2,$$

$$f_n(z', \ln \lambda) = \sum_{m=1}^n \int_{-1}^1 b_m |z' - u|^m \bar{g}_{n-m}(u, \ln \lambda) \mathrm{d}u.$$

Introducing the parameter $\beta = -b_0 + \ln \lambda$ as well as the following transforms:

3

$$\begin{aligned} z' &= e^{-\beta} w, \\ u &= e^{-\beta} \tau, \\ \bar{g}_n(u, \ln \lambda) &= G_n(\tau, \ln \lambda), \\ f_n(z', \ln \lambda) &= e^{-\beta} F_n(w, \ln \lambda), \end{aligned}$$

we obtain the following recurrent chain of singular integral equations defined for all $w$ in the interval $[-e^{-\beta}, e^{-\beta}]$:

$$\int_{-e^\beta}^{e^\beta} G_n(\tau, \ln \lambda) \ln |w - \tau| \mathrm{d}u = F_n(w, \ln \lambda) \tag{22}$$

that we recognize as the Carleman integral equation used in various problems of hydrodynamics, elasticity, etc. Its solution is given by:

$$\begin{aligned} G_n(\tau, \ln \lambda) &= \frac{1}{\pi^2 \sqrt{(\tau + e^\beta)(e^\beta - \tau)}} \\ &\times \left\{ \int_{-e^\beta}^{e^\beta} \frac{\sqrt{(s + e^\beta)(e^\beta - s)} \frac{\partial F_n}{\partial w}|_{w=s}}{s - \tau} \mathrm{d}s + \frac{1}{\ln(2e^\beta/4)} \int_{-e^\beta}^{e^\beta} \frac{F_n(s, \ln \lambda)}{\sqrt{(s + e^\beta)(e^\beta - s)}} \mathrm{d}s \right\}. \end{aligned} \tag{23}$$

We can find that the first terms in the solution of the original equations (16)-(17) are

$$\begin{aligned} \bar{g}(z') &= \frac{2}{\lambda \pi (1 - 2\gamma_e - \ln\left(\frac{\lambda}{4}\right)) \sqrt{1 - z'^2}} \\ &+ \lambda \frac{(7 + 6\gamma_e - 6 \ln 2)((2z'^2 - 1)(2\gamma_e + \ln \frac{\lambda}{4}) - 2z'^2)}{18 \pi \sqrt{1 - z'^2} \left(1 - 2\gamma_e - \ln\left(\frac{\lambda}{4}\right)\right)^2} \\ &+ \mathcal{O}(\lambda^3). \end{aligned} \tag{24}$$

We only provide here the first two terms as higher-order terms rapidly become rather lengthy expressions but we note that the exponent of the divergence of the flux near the edge of the sleeve is preserved in the development. We can now find the function $D(k)$ using (14):

$$\begin{aligned} D(k) &= \frac{J_0(\lambda k)}{1 - 2\gamma_e - \ln\left(\frac{\lambda}{4}\right)} \\ &- \lambda^2 \frac{(7 + 6\gamma_e - 6\ln 2)(J_0(\lambda k) \frac{2 - 2\gamma_e - \ln\left(\frac{\lambda}{4}\right)}{1 - 2\gamma_e - \ln\left(\frac{\lambda}{4}\right)} - \frac{2 J_1(\lambda k)}{\lambda k})}{36 \left(1 - 2\gamma_e - \ln\left(\frac{\lambda}{4}\right)\right)} \\ &+ \mathcal{O}(\lambda^3). \end{aligned} \tag{25}$$

The quantity of interest being the total flux integrated over the surface of the sleeve, let us first recast the flux $j_{\text{small}}(z)$ per unit of surface of the sleeve in physical units,

$$j_{\text{small}}(z) = \frac{\mathcal{D}(c_{\text{sat}} - c_\infty)}{a} \int_0^\infty D(k) \cos\left(k \frac{z}{a}\right) \mathrm{d}k, \tag{26}$$

which becomes, after integration (we only provide here the integration of the first term in the expansion of $D(k)$):

4

$$j_{\text{small}}(z) = \frac{\mathcal{D}(c_{\text{sat}} - c_\infty)}{(1 - 2\gamma_e - \ln\left(\frac{\lambda}{4}\right))\sqrt{\lambda^2 a^2 - z^2}}. \tag{27}$$

Now the total flux $Q_{\text{small}}$ is found by integrating the flux per unit surface over the surface of the sleeve

$$Q_{\text{small}} = \int_{-\lambda a}^{\lambda a} dz \int_0^{2\pi} a d\theta j_{\text{small}}(z), \tag{28}$$

and we find immediately that

$$Q_{\text{small}} = \frac{2D(c_{\text{sat}} - c_\infty)a\pi^2}{1 - 2\gamma_e - \ln\left(\frac{\lambda}{4}\right)}. \tag{29}$$

Table 1: Values of the first $b_n$'s coefficients in equation (19).

| $b_0$ | $1 - 2\gamma_e + \ln 2$ |
|---|---|
| $b_1$ | 0 |
| $b_2$ | $\frac{7 + 6\gamma_e - 6\ln 2}{36}$ |
| $b_3$ | 0 |
| $b_4$ | $\frac{-21 - 20\gamma_e + 20\ln 2}{2400}$ |

## 1.4 Large sleeve $\lambda \gg 1$

The asymptotic result for the total diffusive flux in the limit of large sleeves $\lambda \gg 1$ can be found in an approximate fashion. Let us first introduce the scaled variables $u' = \lambda u$ and $z'' = \lambda z'$ and rewrite (16) as:

$$\int_0^1 \bar{g}(u'/\lambda)\mathcal{L}(z'' - u')du' + \int_1^\lambda \bar{g}(u'/\lambda)\mathcal{L}(z'' - u')du' = 1 \text{ for } |z''| < \lambda \tag{30}$$

Noting that the kernel $\mathcal{L}(x)$ has the following expansions at small and large $x$:

$$\mathcal{L}(x) \approx 1 - 2\gamma_e - \ln\frac{|x|}{2} \text{ for } x \to 0$$

$$\mathcal{L}(x) \approx \frac{\pi}{2|x|} \text{ for } x \to \infty,$$

we now replace the kernel $\mathcal{L}(x))$ appearing in the first (resp. second) integral above by its small x (resp. large x) approximation to obtain:

$$\int_0^1 \bar{g}(u'/\lambda)\left(1 - 2\gamma_e - \ln\left(\frac{|z'' - u'|}{2}\right)\right)du' + \int_1^\lambda \frac{\pi\bar{g}(u'/\lambda)}{2|z'' - u'|}du' = 1 \text{ for } |z''| < \lambda. \tag{31}$$

In the limit of an infinitely long sleeve, we expect the evaporative flux to become invariant by translation along the axis of the fiber. In order to find this constant value, we set $z'' = 0$ in (31) and we look for a constant solution for $\bar{g}$, which simply reads after integration

$$\bar{g}(z') = \frac{1}{2 - 2\gamma_e + \ln 2 + \frac{\pi}{2}\ln\lambda}. \tag{32}$$

Inserting this solution into (14), we find the function $D(k)$ in the limit of large sleeves:

$$D(k) = \frac{\lambda \operatorname{sinc}(\lambda k)}{2 - 2\gamma_e + \ln 2 + \frac{\pi}{2} \ln \lambda}, \tag{33}$$

where $\operatorname{sinc}(x) = \sin(x)/x$ is the cardinal sine function. This leads to the following expression for the constant evaporation flux:

$$j_{\text{large}}(z) = \frac{\mathcal{D}(c_{\text{sat}} - c_\infty)\pi}{2a\left(2 - 2\gamma_e + \ln 2 + \frac{\pi}{2} \ln \lambda\right)}. \tag{34}$$

Integrating the flux per unit surface over the surface of the sleeve, we finally find the total flux in the large sleeve limit:

$$Q_{\text{large}} = \frac{2\lambda \mathcal{D}(c_{\text{sat}} - c_\infty)a\pi^2}{2 - 2\gamma_e + \ln 2 + \frac{\pi}{2} \ln \lambda}. \tag{35}$$

## 1.5 Summary

To summarize, we have the two following limits for the total diffusive flux:

$$Q = \begin{cases} \frac{2\mathcal{D}(c_{\text{sat}} - c_\infty)a\pi^2}{1 - 2\gamma_e - \ln\left(\frac{\lambda}{4}\right)}, & \lambda \ll 1 \\ \frac{2\lambda \mathcal{D}(c_{\text{sat}} - c_\infty)a\pi^2}{2 - 2\gamma_e + \ln 2 + \frac{\pi}{2} \ln \lambda}, & \lambda \gg 1. \end{cases} \tag{36}$$

As a final note, let just add that the results above are valid in the limit where the thickness of the diffusive layer is much larger than any other length scale of the problem (namely $a$ and $\lambda a$). If this thickness becomes comparable to any of those lengths, we expect a saturation of the diffusive flux and the emergence of a different scaling involving this new length.

## 1.6 Comparison between analytical description and numerical results

Figure 1 shows numerical results of the local evaporative flux of sleeves of radius $a = 125$ µm and various lengths. In black lines are plotted the predictions obtained by the model developed here, equation 27 for small aspect ratio sleeves (solid line) and equation 34 for large aspect ratio sleeves (dashed line).

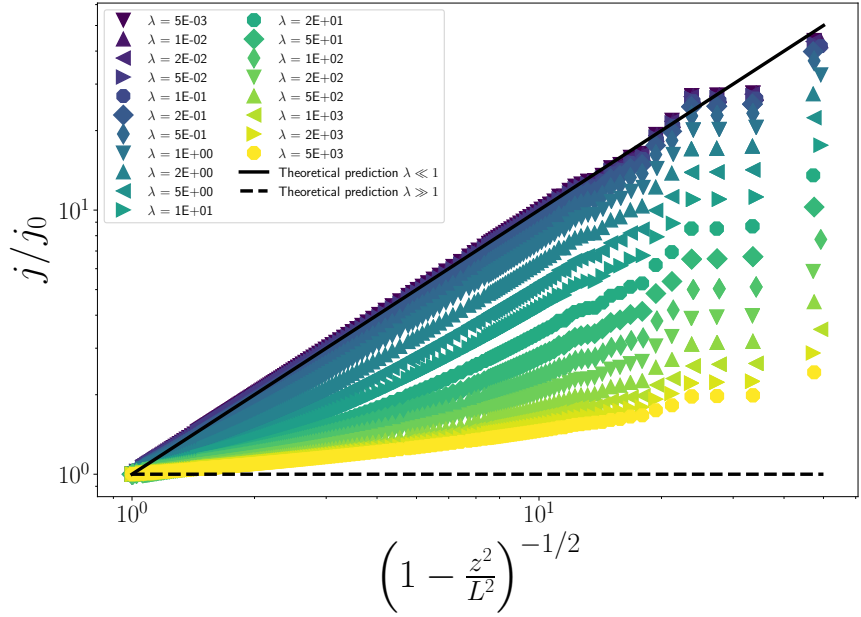

Figure 1: Dimensionless flux density $j/j_0$ as a function of $\left(1-\frac{z^2}{L^2}\right)^{-1/2}$ obtained from numerical computations for $a = 125$ µm. The black solid line corresponds to the local flux described by equation (27) made dimensionless by its value at the center of the fiber ($z = 0$). The black dashed line corresponds to $j = j_0$ as predicted for infinitely long sleeve (Eq. (34)).

## 2 Evaporation of axisymmetric drops on fiber

### 2.1 Comparison between $j_0$ and $j_{\text{sphere}}$

In the figure 2a, we show comparison between the local evaporative flux of barrel-shaped droplets on fiber of radius $a = 125$ µm and local evaporative flux of a spherical drop ($j_{\text{sphere}}$ in black dashed line). The volume of the drops is $\Omega = 1$ µL. In the figure 2b, we plot $j/j_{\text{sphere}}$ as a function of $\Omega^{-1/3}$ for three different contact angles. The spherical case is in black dashed line.

### 2.2 Fitting expression for $j = f(z)$

Figure 3 shows the dimensionless local flux as a function of the dimensionless position along the interface for barrel-shaped droplets on fiber. The volume of the drops is $\Omega = 1$ µL and the fiber radius is $a = 125$ µm. We propose to fit curves of the figure 3 (black dashed lines) with :

$$j(z) = j_0 \left[\beta \left(1 - \frac{z^2}{L^2}\right)^{-\alpha} + (1-\beta)\right], \tag{37}$$

where $\alpha$ and $\beta$ are positive adjustable parameters.

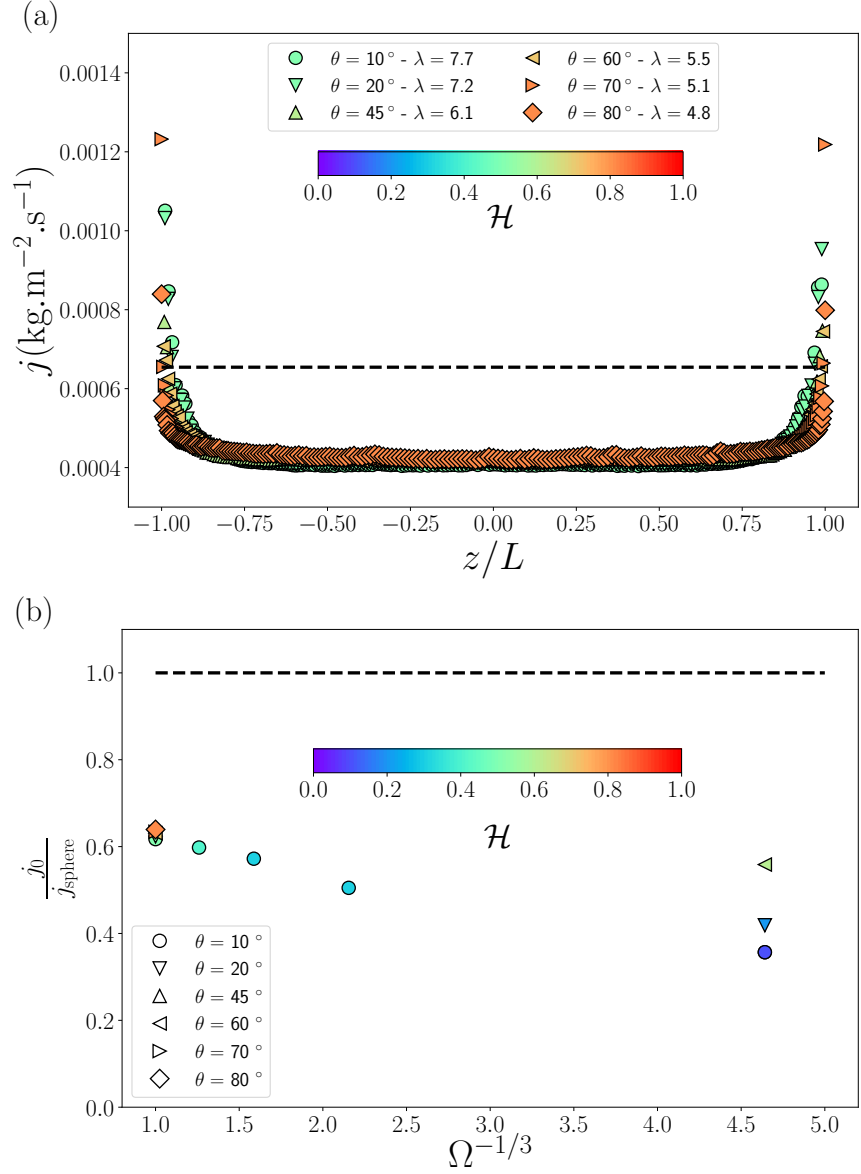

Figure 2: (a) Local flux $j$ as a function of dimensionless position along the air/liquid interface $z/L$ for barrel-shaped droplets on fibers, $a = 125$ µm, $\Omega = 1$ µL and $\theta$ is varied between 10 and 80°. (b) Local flux at the center of the drop $j_0$ divided by local flux of a sphere evaporating under the same experimental conditions as a function of $\Omega^{-1/3}$. Black dashed lines represent local flux of a sphere $j_{\text{sphere}} = \left(\frac{4\pi}{3}\right)^{1/3} \mathcal{D} \left(c_{\text{sat}} - c_\infty\right)\Omega^{-1/3}$. Colors represent the aspect ratio $\mathcal{H} = h_0/L$ of barrel-shaped drops.

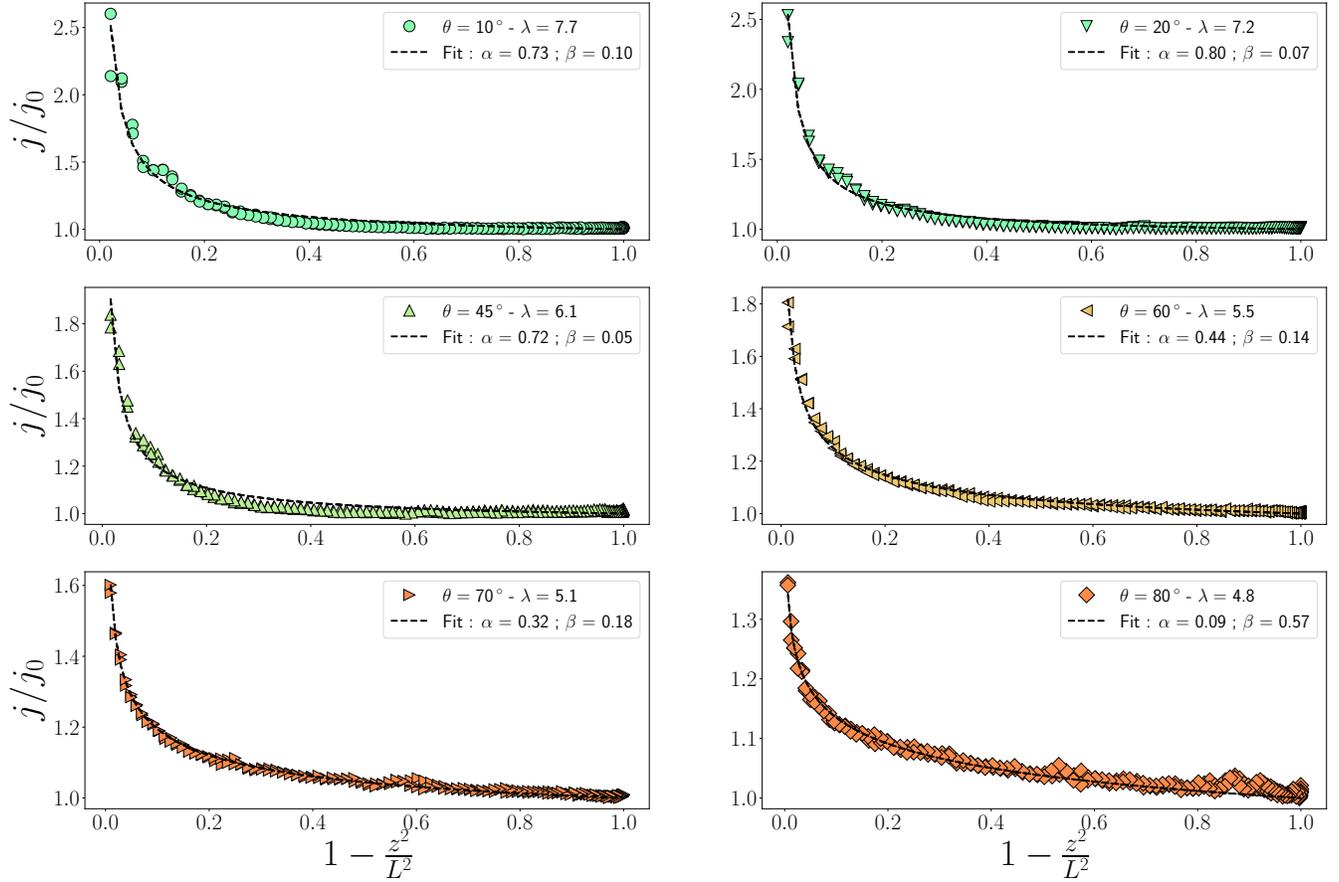

Figure 3: Dimensionless flux density $j/j_0$ as a function of $1 - (z/L)^2$. The points are obtained from numerical computation for drops of volume 1 µL placed on a fiber of radius $a = 125$ µm and various contact angles (see captions). The results are fitted by equation (37) (black dashed line) where $\alpha$ and $\beta$ are given in the caption for each contact angle.

9



\end{document}